\documentclass[sigconf]{acmart}
\usepackage{tikz}
\usepackage{threeparttable}
\usepackage{amsmath}
\usepackage{enumitem}
\usepackage{tcolorbox}
\usepackage{hyperref}
\usepackage{booktabs}
\usepackage{amsmath,amssymb,amsfonts}
\usepackage{algorithmic}
\usepackage{graphicx}
\usepackage{textcomp}
\usepackage{xcolor}
\usepackage[linesnumbered,ruled,vlined,norelsize]{algorithm2e}
\usepackage{tabularx}
\newcommand{\sz}[1]{\textcolor{blue}{[shuzheng: #1]}}
\usepackage{xcolor}

\newcommand{\spara}[1]{\vspace{2mm}\noindent\textbf{#1.}}
\newtheorem{definition}{Definition}

\setcopyright{acmlicensed}
\copyrightyear{2025}
\acmYear{2025}
\acmDOI{XXXXXXX.XXXXXXX}

\acmConference[xxx]{the ACM xxx Conference 2026}{April 28-- 02, 2025}{xxx}

\acmISBN{978-1-4503-XXXX-X/18/06}

\begin{document}

\title{TraceLLM: Security Diagnosis Through Traces and Smart Contracts in Ethereum}

\author{Shuzheng Wang}
\email{swang032@connect.hkust-gz.edu.cn}
\affiliation{%
  \institution{The Hong Kong University of Science and Technology (Guangzhou)}
  \city{Guangzhou}
  \country{China}
}

\author{Yue Huang}
\email{yhuang797@connect.hkust-gz.edu.cn}
\affiliation{%
  \institution{The Hong Kong University of Science and Technology (Guangzhou)}
  \city{Guangzhou}
  \country{China}
}

\author{Zhuoer Xu}
\email{zxu247@connect.hkust-gz.edu.cn>}
\affiliation{%
  \institution{The Hong Kong University of Science and Technology (Guangzhou)}
  \city{Guangzhou}
  \country{China}
}

\author{Yuming Huang}
\email{huangyuming@u.nus.edu}
\affiliation{%
  \institution{National University of Singapore}
  \country{Singapore}
}

\author{Jing Tang}
\authornote{Corresponding author: Jing TANG.}
\email{jingtang@ust.hk}
\affiliation{%
  \institution{The Hong Kong University of Science and Technology (Guangzhou)}
  \city{Guangzhou}
  \country{China}
}

\renewcommand{\shortauthors}{Shuzheng Wang, Yue Huang, Zhuoer Xu, Yuming Huang and Jing Tang}

\begin{abstract}
Ethereum smart contracts hold tens of billions of USD in DeFi and NFTs, yet comprehensive security analysis remains difficult due to unverified code, proxy-based architectures, and the reliance on manual inspection of complex execution traces. Existing approaches fall into two main categories: anomaly transaction detection, which flags suspicious transactions but offers limited insight into specific attack strategies hidden in execution traces inside transactions, and code vulnerability detection, which cannot analyze unverified contracts and struggles to show how identified flaws are exploited in real incidents. As a result, analysts must still manually align transaction traces with contract code to reconstruct attack scenarios and conduct forensics. To address this gap, TraceLLM is proposed as a framework that leverages LLMs to integrate execution trace-level detection with decompiled contract code. We introduce a new anomaly execution path identification algorithm and an LLM-refined decompile tool to identify vulnerable functions and provide explicit attack paths to LLM. TraceLLM establishes the first benchmark for joint trace and contract code-driven security analysis. For comparison, proxy baselines are created by jointly transmitting the results of three representative code analysis along with raw traces to LLM. TraceLLM identifies attacker and victim addresses with 85.19\% precision and produces automated reports with 70.37\% factual precision across 27 cases with ground truth expert reports, achieving 25.93\% higher accuracy than the best baseline. Moreover, across 148 real-world Ethereum incidents, TraceLLM automatically generates reports with 66.22\% expert-verified accuracy, demonstrating strong generalizability. 

\end{abstract}


\maketitle

\section{Introduction}\label{sec:intro}
Ethereum, the second-largest blockchain by market value, extends beyond Bitcoin by supporting Turing-complete smart contracts that automate arbitrary user-defined logic~\cite{wood2014ethereum}. These contracts form the foundation of DeFi~\cite{werner2022sok}, NFTs~\cite{yang2023definition}, and a wide range of decentralized applications. According to Etherscan, more than 78 million smart contracts have been deployed on Ethereum mainnet, with total value locked exceeding 63 billion USD~\cite{EtherscanStat,DefillamaTVL}. Typically, contract logic is written in Solidity, compiled into Ethereum Virtual Machine (EVM) bytecode, and deployed to a dedicated address via transactions~\cite{ethereum2025evm}. Each invocation in transactions is executed step by step by the EVM, producing an execution trace that records low-level operations, message calls, and state transitions. These traces provide the most fine-grained evidence of contract behavior and are central to auditing and post-event analysis~\cite{ethereum2025introduction,zhang2020txspector}.

Despite their importance, execution traces remain difficult to leverage effectively due to their complexity and lack of systematic tooling. This limitation is especially critical for post-incident analysis, as the Ethereum ecosystem continues to face frequent and severe security incidents~\cite{so2021smartest, shou2024llm4fuzz}. In the past two years alone, 218 attacks have been reported on DeFi protocols, with cumulative losses surpassing 953 million USD~\cite{slowmist}. To analyze such incidents, prior research has developed two main lines of work: transaction anomaly detection and code vulnerability detection. The former explored clustering and rule-based methods for labeling suspicious addresses and transactions involved in specific attacks such as reentrancy or phishing~\cite{su2021evil, wu2023defiranger, torres2019art, yuan2020detecting}. However, they offer limited insight into attack strategies hidden within detailed traces of transactions. While some rule-based approaches attempt trace anomaly detection for specific attack types, no comprehensive method systematically analyzes arbitrary traces~\cite{zhang2020txspector}. The latter highlights potential contract flaws through statistical analysis, symbolic execution, fuzzing, and large language model (LLM), yet often fails to analyze unverified contracts without source code and rarely demonstrates how vulnerabilities manifest in real-world exploits. Moreover, even when anomaly transactions or code vulnerabilities are identified, the results are seldom presented in a structured, human-readable form. This underscores a broader gap: the absence of an automated framework that connects anomalies in execution traces with interpretations of contract code flaws and automatically generates comprehensive human-readable security reports.

In this paper, we present TraceLLM, a novel framework that leverages LLM to bridge on-chain execution traces and contract code, thereby enabling human-readable security analysis. Unlike prior approaches that stop at transaction anomaly detection, the framework dives into the trace anomaly detection inside transactions and augments them with contract code, exposing vulnerable functions and explicit attack paths. Our framework allows LLM to use the ability of code understanding, logical reasoning, and multi-source information integration to infer attacker/victim address, attack methods and contract vulnerabilities.

To realize this, we design a modular pipeline comprising four components: Parser, Detector, Extractor, and Analyzer. Parser and Detector normalize user input and collect all relevant transactions and contract information, while addressing common challenges such as proxy resolution and creator detection. Extractor combines the traditional decompile tool with LLM-based refinement to reconstruct contract code even in the absence of verified source code. Analyzer uses numerical and semantic features from traces to detect anomaly traces, and then integrates decompiled code to detect attack mechanism and generate structured, human-readable incident reports. Through extensive empirical evaluation on real-world incidents, TraceLLM demonstrates robust performance in identifying attacker/victim address, uncovering attack execution, and automatically generating detailed security reports. To our knowledge, this is the first approach that establishes a reproducible benchmark for anomaly trace detection and automated report generation in blockchain security.

\spara{Contributions}

In summary, our main contributions are as follows:
\begin{itemize}
    \item We propose TraceLLM, the first LLM-powered automated blockchain security analysis framework. TraceLLM derives the human-readable report from execution traces and contract code, enabling automatic identification of attacker/victim addresses and underlying attack mechanism.
    \item We propose a method to automatically extract anomalous execution paths from raw traces and construct the first anomaly trace dataset containing 11{,}228 execution paths, where our method identifies 83.92\% of anomaly execution paths.
    \item To tackle the prevalent issues of missing source code in real-world smart contracts, we design an enhanced decompile module that improves decompilation precision by 8.52\% over the widely used Etherscan decompiler. 
    \item We manually collect and curate a blockchain security incident dataset and design pipelines for multiple code analysis schemes to automatically generate security reports, forming proxy baselines for systematic comparison. On 27 real-world incident cases, TraceLLM achieves 85.19\% precision in attacker/victim identification and 70.37\% factual precision in security reports, significantly outperforming other representative tools. To evaluate generalizability, we generate security reports for 148 Ethereum incidents, with an expert-verified average precision of 82.43\% in attacker/victim identification and 66.22\% in overall reports.
\end{itemize}

\section{Backgrounds}

\subsection{Large Language Models}
Current mainstream Large Language Models (LLMs), such as GPT~\cite{ouyang2022training}, Deepseek~\cite{bi2024deepseek} and LLaMA~\cite{touvron2023llama}, are primarily built on the Transformer architecture. Trained on massive text corpora, these models demonstrate strong capabilities in both language understanding and generation. Their applications extend into a wide range of domains: in blockchain security area, LLMs are increasingly adopted for vulnerability detection~\cite{yu2025smart}, automated contract generation~\cite{peng2025soleval}, code auditing~\cite{ma2025combining} and program analysis~\cite{ma2025opdiffer}, while also assisting in vulnerability repair~\cite{wang2024contracttinker} and software testing~\cite{sun2025adversarial}. Moreover, LLMs can be extended through retrieval-augmented generation (RAG)~\cite{li2025scalm}, domain-specific fine-tuning~\cite{hu2024zipzap}, and parameter-efficient adaptation techniques~\cite{wangkasa}, which further enhance their applicability and reliability in specialized contexts.

\begin{figure*}[!bpt]
    \centering
    \includegraphics[width=0.98\textwidth]{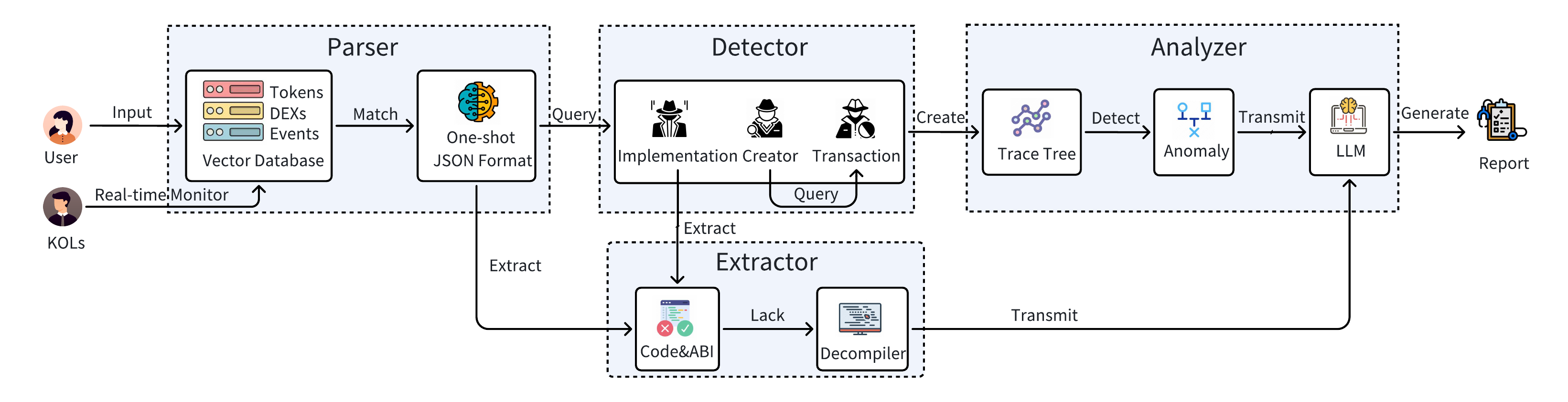}
    \caption{A high-level workflow of TraceLLM.}
    \label{fig:workflow}
\end{figure*}

\subsection{Ethereum Smart contracts}
% EVM运行 Trace Transaction
Ethereum enables the deployment of smart contracts, self-executing programs that encode business logic directly on the blockchain. These contracts allow decentralized applications to operate without intermediaries, automatically enforcing rules and agreements. Smart contracts are written in high-level languages such as Solidity and then compiled into EVM bytecode, which can be executed by all Ethereum nodes in a deterministic manner~\cite{ethereum2025evm}.

The execution of smart contract code within the EVM involves a stack-based architecture where instructions manipulate a finite stack, memory, and persistent storage~\cite{ethereum2025introduction}. Every operation in the EVM consumes “gas,” a unit representing computational cost, to prevent infinite loops and incentivize efficient computation. Transactions trigger the EVM to interpret the compiled bytecode of the contract, step by step, modifying the global state according to the logic defined by the developer. During execution, each opcode updates the EVM state, including stack contents, memory, storage, and the program counter, providing a precise, reproducible model of contract behavior.

Trace is an essential aspect of understanding and analyzing EVM execution. A trace records the step-by-step execution of contract instructions, capturing state transitions, opcode execution, and gas usage. Tracing allows developers and researchers to audit contract behavior, debug logic errors, and detect vulnerabilities such as reentrancy or integer overflows. By examining traces, one can reconstruct the exact sequence of operations performed by a contract, offering insights into how smart contracts interact with one another and with the Ethereum state. Consequently, EVM execution traces form the foundation for formal verification, security analysis, and performance optimization of smart contracts.

% 代码 运行

\section{TraceLLM Overview}\label{sec:Overview}

In this section, we present the overall design of TraceLLM, a modular framework for security analysis of Ethereum smart contracts. At a high level, TraceLLM accepts either natural language queries from end users or monitoring signals from key opinion leaders (KoLs), such as security researchers and watchdog accounts. These inputs are converted into structured analysis tasks, which are subsequently enriched with on-chain execution traces and decomiled contract code. Ultimately, TraceLLM produces comprehensive behavioral analysis reports that reveal anomalous patterns, identify attacker–victim relations, and explain underlying vulnerabilities.

As illustrated in~\autoref{fig:workflow}, TraceLLM operates through four sequential stages. First, the Parser employs a retrieval-augmented generation (RAG) system to map unstructured incident descriptions or informal alerts to precise analysis scopes, namely address sets and block intervals. In parallel, KoLs provide high-signal external intelligence by flagging suspicious large-value transfers or suspected exploit transactions. Second, the Detector continuously tracks transactions of target addresses, logging invoked methods, transferred values, and links to associated logic contracts and suspicious contracts. Third, the Extractor enriches these raw traces by retrieving source code and ABIs from blockchain explorers; when unavailable, bytecode decompilation is applied to recover approximate program structure. Finally, the Analyzer reconstructs the trace-level execution tree and detects anomaly trace paths. LLMs are then employed to fuse execution traces with contract code, generating high-level interpretations of abnormal behaviors.

The detailed workflow of TraceLLM is presented across~\autoref{sec:ParserDetector} and~\autoref{sec:analyzerExtractor}. In~\autoref{sec:ParserDetector}, the Parser and Detector modules are described. This section explains how user inputs and KoL signals are parsed and transformed into generalized actionable scopes, followed by the mechanism for detecting the logic contract, the contract creator, and related transactions. Subsequently,~\autoref{sec:analyzerExtractor} presents the Extractor and Analyzer modules. It details the retrieval of source code and the decompilation of contract bytecode. Trace-level anomaly execution paths are then detected, and these enriched results are synthesized using LLMs to produce comprehensive analysis reports. Finally, we evaluate our TraceLLM in~\autoref{sec:eval} and discuss the extensibility and limitations in~\autoref{sec:discussion}.

\section{Parser \& Detector Modules}\label{sec:ParserDetector}

To enable rigorous incident analysis, TraceLLM first translates vague user inputs into machine-readable on-chain data. The Parser resolves natural-language descriptions into concrete addresses and temporal scopes, while the Detector enriches this scope by uncovering proxy implementations, contract creators, and all relevant transactions. Together, these modules establish a precise analysis target that grounds subsequent extraction and reasoning.

\subsection{Parser}
The Parser functions as the entry point of TraceLLM, transforming heterogeneous and often unstructured information into standardized analysis scopes. Beyond handling natural-language queries and incident descriptions, it continuously ingests external alerts from trusted sources such as KoLs and reporting platforms, thereby capturing emerging events in real time. Through a retrieval-augmented pipeline combined with a one-shot LLM normalization stage, inputs ranging from explicit contract identifiers to vague textual references are mapped to verifiable on-chain entities and converted into machine-readable addresses and block ranges, establishing a reliable foundation for downstream analysis. 

Through the Parser, descriptive queries are resolved into concrete contract sets and temporal scopes. We maintain a domain-specific knowledge base, organized into semantically coherent units such as tokens, DEX pools, and historical security incidents. Token information is sourced from Trust Wallet~\cite{trust-wallet}, which provides a comprehensive and up-to-date collection of data for thousands of crypto tokens. For DEX information, a script is implemented to extract pool addresses from major Ethereum-based DEX, including Uniswap V2\&V3, SushiSwap, and Curve. For security events, hacking incidents on Ethereum mainnet since 2023 are manually collected from DeFiHackLabs~\cite{DeFiHackLabs}. Each unit is embedded into a vector space to enable semantic similarity search, facilitating robust mapping from human-readable descriptions to canonical on-chain address identifiers. Given a query, candidate entities are first extracted. RAG is then applied to ground the LLM with the most relevant entries, thereby reducing hallucination and improving resolution fidelity. In parallel, human-readable temporal expressions are normalized into precise block intervals.

In addition, the Parser continuously monitors signals from security-focused accounts and KoLs on X; new alerts are funneled into the same retrieval-augmented pipeline, yielding consistent representations for both user-driven and externally observed events. After retrieval, a one-shot prompt is invoked that integrates the original query with the retrieved context to produce a strictly structured JSON object specifying the contract list and block range. This in-context design enforces determinism, thereby defining a definitive machine-readable input for the Detector module. The prompt template is illustrated in~\autoref{fig:one_shot_prompt} in Appendix.

\iffalse

\begin{figure}[!bpt]
    \centering
    \includegraphics[width=0.47\textwidth]{figures/llm4refine.pdf}
    \caption{LLM-Refined code extraction framework.}
    \label{fig:llm4refine}
\end{figure}

To extract the code, we begin by checking if the source code of the target contract is available via the Etherscan API~\footnote{https://etherscan.io/}. If the request returns the source code, it would be forwarded to the subsequent step. However, many Ethereum smart contracts lack publicly available source code, particularly attack contracts deployed by exploiters and hackers. In this section, we introduce our strategy to accurately extract the code of smart contracts.

We have implemented an LLM-Refined code extraction framework to ensure comprehensive analysis across all scenarios. We utilize the Panaromix decompiler~\footnote{https://github.com/eveem-org/panoramix} to decompile the contract based on its deployed bytecode. Subsequently, following the methodology outlined in~\cite{tan2024llm4decompile}, we utilize Google Gemini 2.0 Flash to refine the decompilation output before delivering it to LLM module for report generation.
\fi

\subsection{Detector}
\label{sec:detector}
The Detector begins operation upon receiving precise address and time range inputs from the Parser. It addresses three key dimensions: the identification of logical contracts behind proxy architectures, the attribution of contracts to their creators, and the collection of execution traces for relevant transactions. These functions are implemented through the Implementation Detector, Creator Detector, and Transaction Detector. Single address and time range inputs are transformed into comprehensive information on contracts and transactions that may be involved in security events, establishing the foundation for subsequent contract code extraction and trace-level anomaly detection.

\subsubsection{Implementation Detector}

To accurately resolve logical contract addresses within Ethereum's proxy architecture, we adapt a streamlined detection framework, leveraging the execution tracing capabilities of a locally deployed Ethereum node~\cite{William2023Proxy}. This approach emphasizes precision and efficiency, eliminating the need for extensive pattern matching or historical state traversal.

\vspace{-1mm}
\spara{Step 1: Proxy contract identification}
For a given on-chain contract address, runtime bytecode is first retrieved using the \texttt{eth\_getCode} RPC method from the local node. The bytecode is subsequently disassembled to detect the presence of the \texttt{DELEGATECALL} opcode, which enables proxy-based invocation. Contracts lacking \texttt{DELEGATECALL} are immediately classified as non-proxy and excluded from further analysis. This static pre-filter minimizes tracing overhead.

\vspace{-1mm}
\spara{Step 2: Logical address resolution via execution trace}
For contracts identified as proxies, controlled execution tracing is performed using the \texttt{debug\_traceCall} RPC method of the local node. A call is issued with a random, non-matching function selector to ensure execution passes through the \texttt{fallback} function. During tracing, execution steps are monitored for the \texttt{DELEGATECALL} instruction, and the target address is extracted directly from the EVM stack. This target corresponds to the current logical contract address in use. For minimal proxies conforming to EIP-1167, the address is hard-coded in the bytecode and is recovered directly. For storage-based proxy patterns, the address is retrieved from the storage slot accessed immediately prior to the \texttt{DELEGATECALL}, obviating the need for pre-defined slot mappings.

This trace-based method operates through a fixed two-step procedure and does not depend on standard storage keys, ABI-level methods such as \texttt{implementation()}, nor on verified source code. By combining static bytecode inspection with dynamic execution tracing, high accuracy is achieved across diverse proxy patterns while maintaining low computational complexity. This makes the approach particularly suitable for large-scale empirical studies, where resolving the current logical contract address is a prerequisite for downstream analysis.

\subsubsection{Creator Detector}

The Creator Detector resolves the creator of a given contract and enumerates other contracts deployed by the same address within the same block range. This expands the analysis from a single suspicious contract to the broader activity scope of its creator, enabling comprehensive threat detection.

Given a contract address, the deployment transaction is identified as the earliest transaction in which the address appears in the transaction receipt, with \texttt{to} set to \texttt{null}. The \texttt{from} field of this transaction is recorded as the creator address. Indexed blockchain explorers can be leveraged to directly obtain the creator and deployment transaction hash~\cite{etherscanCreator}. If the creator itself is a contract (e.g., a factory contract), recursive resolution is applied to trace back to the originating externally owned account.

Once the creator address is resolved, all contracts deployed by it are enumerated using the local node. Blocks within the specified time range are iterated, and transactions with \texttt{from} equal to the creator address and \texttt{to} set to \texttt{null} are extracted. For each contract creation transaction, the contract address is retrieved from the transaction receipt and added to the creator’s deployment set. By combining creator identification with local full-node enumeration, the Creator Detector provides a comprehensive view of a creator's deployment history within the analysis window. This enables the linkage of multiple suspicious contracts to a single actor, uncovers large-scale malicious deployments, and supports thorough analysis of the security event.

\subsubsection{Transaction Detector}
\label{sec:trace}
Building on the resolution of logical contracts and their creators, the Transaction Detector is responsible for collecting the transactions associated with the identified addresses. For each target address $A$, we query the local node to retrieve all transactions in the specified temporal window to ensure alignment with incident-related activities. Subsequently, we apply the \texttt{debug\_traceTransaction} RPC interface to extract detailed execution traces, capturing low-level call semantics (\texttt{CALL}, \texttt{DELEGATECALL}, \texttt{STATICCALL}, and \texttt{CREATE}), caller–callee relations, transferred values, and call data. To enhance interpretability, the first four bytes of call data are parsed as function selectors and cross-referenced with public signature databases (e.g., the Ethereum Signature Database), allowing recovery of human-readable function names where possible. The resulting output is a temporally ordered sequence of structured trace events that preserves the fidelity of execution semantics and serves as the foundation for subsequent analysis.
\section{Analyzer \& Extractor Modules}
\label{sec:analyzerExtractor}
After the precise addresses and transactions are obtained from~\autoref{sec:ParserDetector}, the collected data will be processed and transmitted to LLM for analysis. The Extractor retrieves the contract code from the address and decompiles undisclosed contracts. The Analyzer restructures the complex traces within the transaction into a call tree and identifies anomaly execution paths. The processed information of traces and contract code is then provided to LLM to generate the final human-readable security reports.

\subsection{Extractor}
\label{sec:extractor}

\begin{figure}[!bpt]
    \centering
    \includegraphics[width=0.9\linewidth]{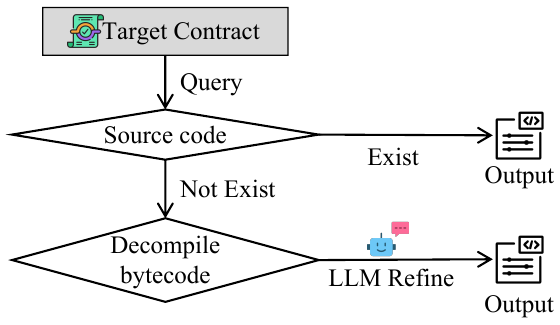}
    \caption{LLM-refined code extractor workflow.}
    \label{fig:llm4refine}
\end{figure}

Meaningful semantic interpretation of contract behavior necessitates understanding its underlying code logic. Accordingly, the Extractor module retrieves contract metadata for all addresses identified by the Detector. Within the Ethereum ecosystem, every deployed smart contract stores its compiled bytecode on-chain for execution by the EVM. Although bytecode suffices for runtime execution, many contracts voluntarily submit source code and Application Binary Interface (ABI) to public verifiers such as Etherscan. Verified source code offers a richer semantic view, enabling deeper security analysis and structured report generation. However, adversarial contracts involved in security incidents are rarely verified, as malicious actors typically withhold source code and ABI to impede reverse engineering and forensic efforts. Consequently, the Extractor must often rely on raw bytecode and decompilation to recover semantic insights into the contract’s logic. To address these challenges, the Extractor operates in two phases:

\vspace{-1mm}
\spara{Step 1: Retrieval of verified metadata}
For each unique contract address identified in the execution trace (\autoref{sec:call_tree}), we query blockchain explorer APIs such as Etherscan to retrieve its verified source code and ABI. When available, the ABI enables precise mapping between 4-byte function selectors observed in the trace and their corresponding human-readable function names, while the source code allows for advanced static analysis, including control-flow reconstruction and vulnerability scanning.

\vspace{-1mm}
\spara{Step 2: Bytecode decompilation}
If no verified source code or ABI is available, we directly obtain the deployed bytecode from the Ethereum network via the local node. This bytecode is then processed using the \textit{Panoramix} decompiler, which translates EVM bytecode into a higher-level pseudocode representation. Although decompilation cannot perfectly recover the original source semantics, it reveals contract functions, control structures, and storage access patterns. These artifacts are often sufficient to identify malicious logic, correlate related contracts deployed by the same actor, and support further static or dynamic analysis. However, the results from Panoramix still lack readability. While recent research has shown that LLMs can further enhance code readability~\cite{tan2024llm4decompile}. Based on this, we use LLM to refine the output results of Panoramix. The prompt for LLM is provided in~\autoref{fig:prompt_refine} in Appendix. By combining verified metadata with decompiled code, the Extractor ensures that subsequent analysis stages have access to the code available for each contract, regardless of its verification status.

\subsection{Analyzer}
\label{sec:analyzer}

The Analyzer operates on enriched transactions with execution traces provided by the Parser and Detector module in~\autoref{sec:ParserDetector}. However, it will consume large tokens for directly passing these traces to LLM, and too many traces often obscure the anomaly execution paths within them. The goal of the Analyzer is to transform these flat execution traces into semantically rich structures, extract behavior-relevant patterns, and identify the anomaly trace paths.

\subsubsection{Reconstruct the Call Tree}
\label{sec:call_tree}

\begin{figure}[!bpt]
    \centering
    \includegraphics[width=1\linewidth]{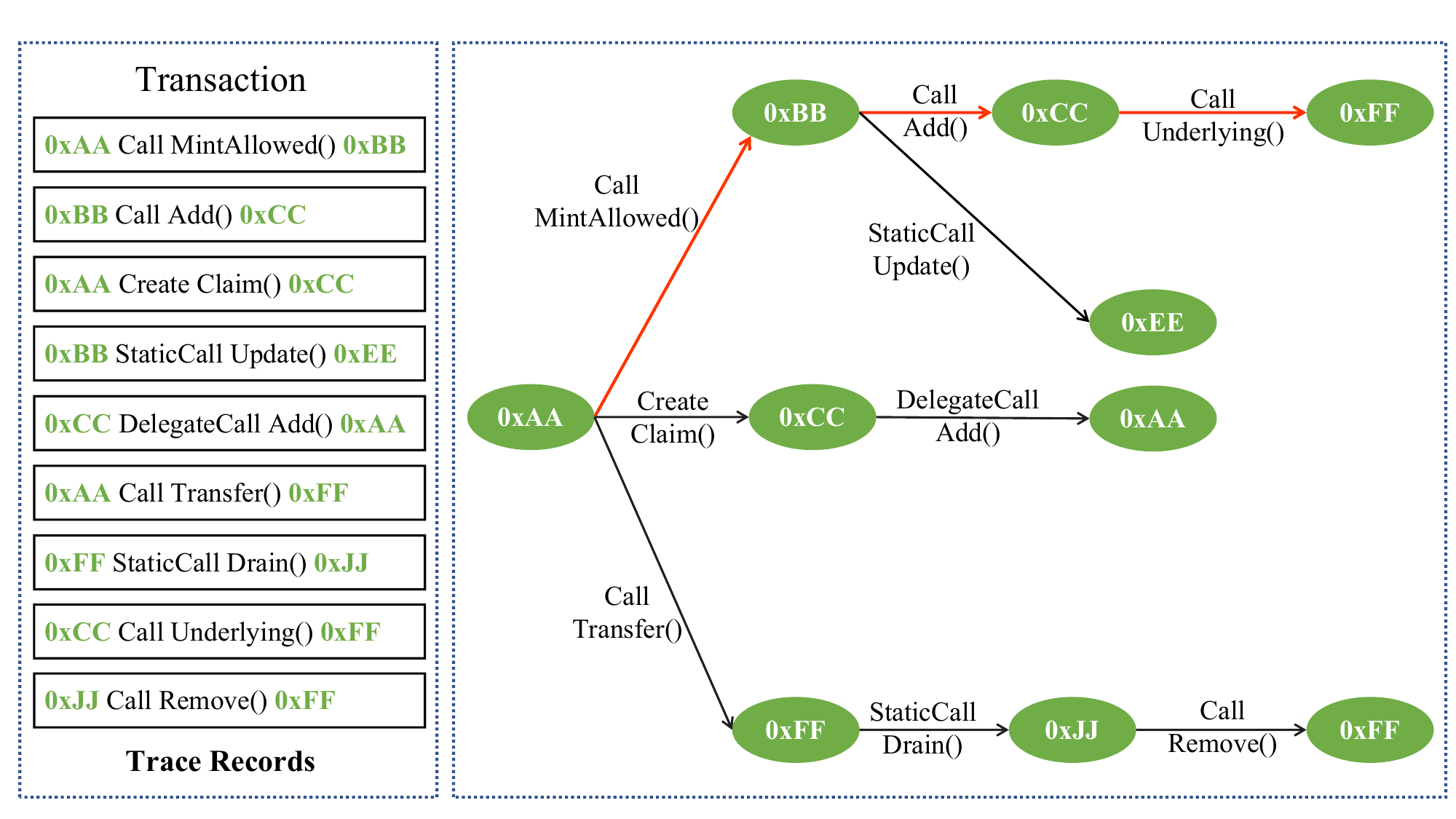}
    \caption{Reconstruction of call trees from flat traces.}
    \label{fig:calltree}
\end{figure}

In the EVM, contract execution may trigger nested invocations of other contracts through low-level opcodes, as mentioned in~\autoref{sec:trace}. These invocations are processed in a strict first-come first-served rule. This mechanism naturally induces a hierarchical structure, with the transaction entry point as the root and deeper layers corresponding to nested calls.

However, the raw transaction trace only records calls as a flat sequence without preserving their hierarchy. To recover the execution structure, we reconstruct the trace into a call tree. In this representation, nodes correspond to externally owned accounts (EOAs) or contracts, while directed edges capture call events between them. Each edge is annotated with metadata including the invoked method, transferred value, call type, and execution result. The resulting call tree thus encodes both the control relationships and semantic attributes of invocations, providing a structured view of the execution flow. This abstraction enables downstream analyses such as identifying suspicious paths, extracting contextual subgraphs, and reasoning about behavior-specific patterns. By bridging the raw transaction detector output with higher-level semantics, the call tree serves as a foundational data structure for understanding transaction behavior in its full execution context.

Formally, given a flat execution trace $\mathcal{T} = \left[\texttt{call}_0,\texttt{call}_1,\ldots,\texttt{call}_n\right]$, each entry in the trace corresponds to a low-level EVM call and is annotated with a tuple of attributes $\texttt{call}_i = \left(\texttt{from}, \texttt{to}, \texttt{method}, \texttt{value}, \texttt{calltype}\right)$. Here, \texttt{from} and \texttt{to} denote the caller and callee addresses (e.g., EOAs or contracts), \texttt{method} records the invoked function signature, \texttt{value} is the amount of ETH transferred in the call, and \texttt{calltype} specifies the low-level opcode used for the invocation (e.g., CALL, DELEGATECALL). We reconstruct the tree structure from this flat list by identifying child and sibling relationships.

\begin{definition}[Rules of call tree reconstruction]
\label{def:call_tree}
The child relationship and sibling relationship can be rebuilt from the EVM trace by the following rules:
\begin{itemize}
    \item \emph{child}: $call_i$ is attached as a child node of $call_{i-1}$ if and only if $call_i.from = call_{i-1}.to$.
    \item \emph{sibling}: If there exists a consecutive sequence of calls \\ $call_i, call_{i+1}, \ldots,call_{i+k}$ such that $call_{i+j}.from = call_i.from$ for all $j \in \{1,\ldots,k\}$, then these calls are considered sibling nodes at the same depth.
\end{itemize}
\end{definition}

\begin{algorithm}[!bpt]
\setlength{\hsize}{0.95\linewidth}
\small
\SetAlFnt{\small}
\SetAlCapFnt{\small}
\SetAlCapNameFnt{\small}
\SetAlgoLined
\SetKw{KwForIn}{in}
\KwIn{Flat trace $\mathcal{T} = [\texttt{call}_0,\dots,\texttt{call}_n]$}
\KwOut{Call trees $\mathcal{F}$}

$\mathcal{F} \gets [~]$\;
$S \gets [~]$\;

\For{$i$ \KwForIn $0 \dots n$}{
    \If{$i=0$ \textbf{or} $\texttt{call}_i.\mathrm{from} \neq \texttt{call}_{i-1}.\mathrm{to}$}{
        $T \gets \text{Tree}(\texttt{call}_i)$\;
        $\mathcal{F}.\text{append}(T)$\;
        $S \gets [\texttt{call}_i]$\;
    }
    \Else{
        \While{$S \neq [~]$ \textbf{and} $\texttt{call}_i.\mathrm{from} \neq \mathrm{top}(S).\mathrm{to}$}{
            $S.\text{pop}()$\;
        }
        $p \gets \mathrm{top}(S)$\;
        $p.\mathrm{children}.\text{append}(\texttt{call}_i)$\;
        $S.\text{push}(\texttt{call}_i)$\;

        \While{$i < n$ \textbf{and} $\texttt{call}_{i+1}.\mathrm{from} = \texttt{call}_i.\mathrm{from}$}{
            $i \gets i+1$\;
            $p.\mathrm{children}.\text{append}(\texttt{call}_i)$\;
            $S.\text{push}(\texttt{call}_i)$\;
        }
    }
}

\KwRet{$\mathcal{F}$}\;
\caption{Reconstruction of call trees from flat EVM traces.}
\label{alg:calltree}
\end{algorithm}

The child relation denotes that the callee of the preceding call assumes the role of the caller in the subsequent call, thereby encoding nested invocation. Sibling calls share a common parent and arise when a contract issues multiple independent calls before returning.

 ~\autoref{alg:calltree} outlines the procedure. A stack is maintained to track the current call context. For each trace entry, either a new tree is initiated (if the caller does not match the stack top) or the entry is attached as a child of the most recent matching parent. Consecutive calls originating from the same caller are grouped as siblings. The resulting call tree $\mathcal{F}$ contains one tree for each independent top-level call within the transaction. This representation explicitly encodes parent–child relations among calls, facilitating reasoning about execution contexts and value propagation in a single transaction, as shown in~\autoref{fig:calltree}.

\iffalse
%\subsection{call tree structure}
We model the call structure of Ethereum transactions as a directed graph $G=(V, E)$ with vertex set $V$ and edge set $E = \{e1, e2, \ldots\} \subseteq V \times V$. Each node $v\in V$ represents a single function-invocation event. A directed edge $(v_i,v_j)\in E$ means that invocation $v_i$ directly triggers invocation $v_j$, i.e.\ $v_i$ invokes $v_j$, which is similar to the contract invocation graph in prior work. Thus the program execution is modeled as a directed invocation graph. We annotate each call-event with attributes $v_i = (from, to, method, value, calltype)$. Here “caller” and “callee” identify the invoking and invoked entities (e.g. contract addresses), “method” is the function signature, “value” is any transferred value, “depth” is the call-stack depth, and “calltype” indicates the type of call (e.g. direct vs. delegate). 

Each attribute function maps $v$ to its value; for example, $\mathit{caller}(v_0)=A$ if address $A$ invoked $v_0$, and $\mathit{depth}(v_0)=2$ if $v_0$ is two calls deep from the transaction origin. Since $G$ is acyclic, one may also implicitly define a root with no caller (or a special external caller address). In summary, $G=(V,E)$ is a directed invocation graph whose edges encode caller-callee relations, and each $v\in V$ carries rich metadata (caller, callee, method, value, depth, calltype) describing that invocation. This graph-based model compactly encodes the execution trace and will be the basis for scoring and subgraph extraction.
\fi

\subsubsection{Anomaly Execution Path Detection}

\begin{table}[t]
\centering
\small
\begin{threeparttable}
\begin{tabularx}{\linewidth}{>{\raggedright\arraybackslash}X p{3.6cm}}
\toprule
\textbf{Method Signatures} & 
\textbf{Vulnerability Class} \\
\midrule

\texttt{selfdestruct()} & Contract Termination\tnote{\textsuperscript{\dag}} \\

\texttt{fallback()}, \texttt{receive()} & Fallback Abuse\tnote{\textsuperscript{\ddag}} \\

\texttt{initialize()} & Re-initialization Flaws\tnote{\textsuperscript{\S}} \\

\texttt{transfer()}, \texttt{transferFrom()} & Silent Transfer Failure\tnote{\textsuperscript{\P}} \\

\texttt{onlyOwner()}, \texttt{hasRole()} & Access Control Misconfig\tnote{\textsuperscript{\textsection}} \\

\texttt{ecrecover()}, \texttt{assert()}, \texttt{require()} & Signature Logic Flaws\tnote{\textsuperscript{$\|$}} \\

\texttt{address.call()}, \texttt{ExternalContract.any()} & Arbitrary External Call\tnote{\textsuperscript{**}} \\

\texttt{tokensReceived()}, \texttt{tokensToSend()} & Reentrancy Callback\tnote{\textsuperscript{\dag\dag}} \\

\texttt{balanceOf()}, \texttt{sweepToken()}, \texttt{drain()} & Unrestricted Withdrawal\tnote{\textsuperscript{\ddag\ddag}} \\

\texttt{isOperationReady()}, \texttt{beforeCall()} & Governance Bypass\tnote{\textsuperscript{\S\S}} \\

\bottomrule
\end{tabularx}
\begin{tablenotes}
\footnotesize
\item[\textsuperscript{\dag}] SWC-106: Unhandled Self-Destruct.
\item[\textsuperscript{\ddag}] SWC-104: Unexpected Ether Receive.
\item[\textsuperscript{\S}] SWC-118: Incorrect Constructor.
\item[\textsuperscript{\P}] SWC-135: Incorrect Return Value Handling.
\item[\textsuperscript{\textsection}] SWC-124: Access Control. 
\item[\textsuperscript{$\|$}] SWC-122: Signature Validation.
\item[\textsuperscript{**}] SWC-112: Delegatecall to Untrusted Contract.
\item[\textsuperscript{\dag\dag}] SWC-107: Reentrancy.
\item[\textsuperscript{\ddag\ddag}] SWC-105: Unrestricted Withdrawals.
\item[\textsuperscript{\S\S}] No official SWC ID, commonly categorized under Governance Exploits.
\end{tablenotes}
\caption{Suspicious method signatures and associated vulnerabilities.}
\label{tab:critical-functions}
\end{threeparttable}
\end{table}

While the reconstructed call tree offers a comprehensive view of execution flow, its size and complexity present significant challenges for analysis. Large-scale security incidents often involve a single malicious transaction triggering extensive internal call cascades, resulting in hundreds of trace entries. For example, the May 28, 2025 attack on Cork Protocol, which caused a loss of 12 million USD, included 362 traces within the attack transaction alone.

Feeding such large call trees directly into downstream LLMs is inefficient, as it consumes excessive input tokens and risks diluting key malicious execution paths within overwhelming context. To address this, a dedicated feature extraction and classification pipeline has been designed to rank and highlight suspicious subpaths prior to LLM-based reasoning, enabling more focused and token-efficient analysis. Rather than inputting the entire tree into reasoning modules, multiple features are extracted from each root-to-leaf path, followed by supervised classification to distinguish adversarial from benign execution paths.

The feature set combines structural–semantic numerical features derived from the call tree with lexical features from the sequence of invoked methods. The structural–semantic features are inspired by the backward and forward causality tracker used in operating system advanced persistent threat detection~\cite{liu2018towards}. It considers four key factors that strongly influence the suspiciousness of a path: path fanout, path frequency, path depth, and method anomaly. Concurrently, term frequency inverse document frequency(TF-IDF) features are extracted from ordered sequences of method signatures along each path, treated as textual tokens. Mathematical definitions of these features are provided below.

As mentioned in~\autoref{sec:call_tree}, we reconstruct a hierarchical structure that explicitly captures the nested nature of contract execution. 
%We define a call tree as a rooted directed tree $\mathcal{G}(\mathcal{V},\mathcal{E},r)$, where $\mathcal{V}$ is the set of nodes, and each nodes denotes an externally owned account (EOA) or a contract address participating in the execution. $\mathcal{E} = \{e_1, e_2, \ldots\,e_m\} \subseteq V \times V$ encodes invocation events: for any edge $e = (u,v)\in \mathcal{E}$ with $u,v \in \mathcal{V}$, $v_p$ is the caller and $v_q$ is the callee. Each edge $e_k$ is annotated with metadata extracted from the execution trace—such as the invoked method signature, transferred value, call type, and execution result—so that a call event can be regarded as the triple $(v_p,v_q;e_k)$. $r \in \mathcal{V}$ denotes the root node of the tree, corresponding to the top-level external transaction. A root-to-leaf path in $\mathcal{G}$ is denoted as $P = \left(v_{p_0}, v_{p_1}, \ldots, v_{p_\ell}\right)$, where $v_{p_0} = r$, $v_{p_l}$ is a leaf node, and $(v_{p_j}, v_{p_j-1}) \in \mathcal{E}, 1 \leq j \leq \ell$.
Let $\mathcal{A}$ be the set of blockchain addresses. We model a single-transaction execution as $\mathcal{G}=(\mathcal{V},\mathcal{E}, r, \mathrm{addr}, \phi)$, where $\mathcal{V}\subseteq\mathcal{A}\times\mathbb{N}$ is the set of address-labeled invocation instances; each $v=(a,k)$ has address $a\in\mathcal{A}$ and occurrence index $k$. The labeling map $\mathrm{addr}:\mathcal{V}\to\mathcal{A}$ returns the underlying address. $\mathcal{E}\subseteq\mathcal{V}\times\mathcal{V}$ encodes parent–child edges; each $e=(u,v)$ denotes a call from $u$ to $v$ with attributes $\phi(e)$ such as method signature, value, call type (e.g., \texttt{CALL}, \texttt{DELEGATECALL}), result, and trace index. We write $\texttt{from}(e)=\mathrm{addr}(u)$ and $\texttt{to}(e)=\mathrm{addr}(v)$. The root $r$ is the top-level invocation. A root-to-leaf path is $P=(v_{p_0},\ldots,v_{p_\ell})$ with edges $e_j=(v_{p_{j-1}},v_{p_j})$. Its edge sequence is $\mathbf{e}(P)=(e_1,\ldots,e_\ell)$; method/value/calltype are accessed via $e_j$.

%To capture the diverse characteristics of attack behaviors, our algorithm considers four key factors that strongly influence the suspiciousness of a path: path fanout, path frequency, path depth, and method anomaly. These factors are computed directly from the reconstructed call tree, ensuring that the scoring process leverages both the topology of the execution and the semantics of individual calls.

\vspace{-1mm}
\spara{Path fanout}
The branching factor quantifies how many distinct downstream calls each node triggers. The raw fanout of $P$ is the sum of out-degrees along the path. It can be drawn as
\begin{equation}
\begin{aligned}
    \deg^+(v)&=|\{u\in V\mid(v,u)\in E\}|,\\ 
    \operatorname{F}&(P)=\sum_{j=0}^\ell\deg^+(v_{p_j}).
\end{aligned}
\end{equation}

Benign transactions, such as token transfers or simple swaps, typically produce narrow and almost linear call patterns, resulting in a low fanout. In contrast, adversarial transactions often trigger a cascade of external calls in rapid succession—for example, interacting with multiple token contracts and liquidity pools to manipulate prices or drain assets.

\vspace{-1mm}
\spara{Path depth}
The nesting depth of a path measures the maximum level of call-stack embedding observed during an execution sequence.  We define the depth of $P$ as
\begin{equation}
\mathrm{D}(P) = \ell + 1,
\end{equation}
where $\ell+1$ is the total number of nodes in $P$. Deep paths indicate reentrancy, recursive creation, or multi-hop manipulations where inner calls perform critical state updates.

\vspace{-1mm}
\spara{Path frequency}
The frequency of a path measures how often a specific call sequence recurs within the same execution trace. We define a path pattern as the ordered sequence of method signatures along the edges. It can be drawn as
\begin{equation}
\label{eq:sig}
\mathrm{sig}(P) = \big[ \texttt{method}(e_1),\, \texttt{method}(e_2),\, \ldots,\, \texttt{method}(e_\ell) \big].
\end{equation}
The frequency of this pattern within the transaction trace is defined as
\begin{equation}
\mathrm{freq}(P) = \Big| \big\{ P' \in \mathsf{Paths}(\mathcal{G}) \,\big|\, \mathrm{sig}(P') = \mathrm{sig}(P) \big\} \Big|,
\end{equation}
where $\mathsf{Paths}(\mathcal{G})$ denotes the set of all root-to-leaf paths in $\mathcal{G}$. 
This metric captures the number of structurally equivalent call paths that exhibit identical functional behavior. Exploit logic often manifests as recurring control patterns, such as repeatedly calling liquidation functions, invoking flash-loan callbacks, or looping over asset operations to maximize impact. 

\vspace{-1mm}
\spara{Path semantic anomaly}
In addition to topological factors, we incorporate semantic signals by quantifying anomalous method invocations along a path. Let $\mathcal{M}$ denote the set of fundamental suspicious method signatures identified through the Smart Contract Weakness Classification (SWC)~\cite{SWC2020}, as summarized in~\autoref{tab:critical-functions}. 
We define the anomaly score as
\begin{equation}
\label{eq:semantic}
\mathrm{S}(P) = \frac{1}{\ell} \sum_{j=1}^\ell \mathbf{1}\big(\texttt{method}(e_j) \in \mathcal{M}\big),
\end{equation}
where $\mathbf{1}(\cdot)$ is the indicator function that evaluates to 1 if the invoked method on edge $e_j$ belongs to $\mathcal{M}$, and 0 otherwise. This factor captures the semantic irregularity of execution by highlighting the density of high-risk methods, such as fallback handlers, \texttt{selfdestruct()} invocations, administrative routines, and reentrancy-sensitive callbacks. 
%Paths exhibiting a high anomaly score are thus more likely to correspond to non-standard execution patterns driven by adversarial intent rather than benign user behavior.

\vspace{-1mm}
\spara{Path TF-IDF representation}
Beyond numerical descriptors of path structure and semantics, we also capture the statistical
salience of method invocations through a term-frequency-inverse-document-frequency (TF-IDF)
representation. We treat each root-to-leaf path $P$ as a `document' whose tokens correspond to the ordered method signatures in $\operatorname{sig}(P)$ defined in Eq.~\ref{eq:sig}. Given the corpus $\mathcal{C}$ of all root-to-leaf paths extracted from the reconstructed call tree $\mathcal{G}$, the term frequency of a method token $t$ in path $P$ is
defined as
\begin{equation}
\mathrm{TF}(t, P) = \frac{\mathrm{count}(t, P)}{\sum_{t' \in \mathrm{sig}(P)} \mathrm{count}(t', P)},
\end{equation}
where $\mathrm{count}(t,P)$ denotes the number of times token t occurs in sig(P). The inverse document frequency is given by:
\begin{equation}
\mathrm{IDF}(t, \mathcal{C}) = \log \frac{|\mathcal{C}|}{1 + |{P' \in \mathcal{C} \mid t \in \mathrm{sig}(P')}|}.
\end{equation}
The TF-IDF weight for token t in path P is then computed as:
\begin{equation}
\mathrm{TFIDF}(t,P) = \mathrm{TF}(t,P) \cdot \mathrm{IDF}(t,\mathcal{C}).
\end{equation}

The resulting TF-IDF vector encodes the relative importance of each method invocation in the context of all observed paths, attenuating the influence of common contract routines while amplifying rare but potentially security-relevant calls. Such rare, high-weight tokens often correspond to functions that appear selectively in exploit logic, e.g., specialized callbacks, privilege-altering routines, or asset-draining primitives, and thus provide an orthogonal semantic signal to the structural–semantic numerical features described above.

\vspace{2mm}
We concatenate the five features described above to form a unified feature vector for each root-to-leaf path. This combined representation is then fed into a binary logistic regression classifier, which outputs the probability that the path corresponds to malicious behavior:
\begin{equation}
Pr\left( y=1 \mid \mathbf{x} \right) = \sigma(\mathbf{w}^{\top} \mathbf{x} + b)
\end{equation}
Here, $\mathbf{x} \in \mathbb{R}^d$ denotes the concatenation of the features, $\mathbf{w}$ and $b$ are model parameters, and $\sigma(\cdot)$ is the sigmoid function. Under this formulation, paths assigned higher probabilities tend to exhibit structural complexity, lexical irregularity, and semantic patterns aligned with high-risk behavior—traits commonly associated with adversarial or exploit-oriented activity. To check the performance of our classifier, we conduct comparative experiments with alternative approaches, including traditional statistical models, machine learning algorithms, deep learning architectures, and graph-based analytical methods in~\autoref{sec:eval}.

\iffalse
Having introduced the four path-level factors—fanout, depth, frequency, and semantic anomaly—we integrate them into a unified scoring function that prioritizes execution paths exhibiting structural complexity and semantic irregularities. To ensure comparability between factors of different scales, we first apply min–max normalization to the fanout and depth metrics across all paths within the same transaction:
\begin{equation}
\widehat{\mathrm{F}}(P)=\frac{\mathrm{F}(P)-\min_{P'}\mathrm{F}(P')}
{\max_{P'}\mathrm{F}(P')-\min_{P'}\mathrm{F}(P')},
\end{equation}

\begin{equation}
\widehat{\mathrm{D}}(P)=\frac{\mathrm{D}(P)-\min_{P'}\mathrm{D}(P')}
{\max_{P'}\mathrm{D}(P')-\min_{P'}\mathrm{D}(P')}.
\end{equation}

The final score assigned to a root-to-leaf path $P$ is then computed as a weighted linear combination of all four factors:
\begin{equation}
\mathrm{Score}(P)=
\alpha \cdot \widehat{\mathrm{F}}(P) + \beta \cdot \widehat{\mathrm{D}}(P) + \gamma \cdot \frac{1}{\mathrm{Fr}(P)} + \delta \cdot \mathrm{S}(P),
\end{equation}
where $\alpha,\beta,\gamma,\delta \geq 0$ are tunable weights reflecting the relative importance of each factor. The inverse frequency term highlights rare behavioral patterns, while the anomaly component emphasizes the semantic presence of security-critical method calls. High-scoring paths under this formulation correspond to execution sequences that are structurally broad, deeply nested, repetitive, and semantically irregular, which together are strong indicators of adversarial or exploit-driven activity.
\fi

\subsubsection{$k$-hop Enclosing Subgraph}
\label{sec:k-hop}

In practice, analysts require not just a ranking of suspicious paths, but also rich context to support accurate root-cause analysis. While scoring-based methods highlight anomalous traces, effective investigation often demands a broader view beyond isolated paths. To improve interpretability and reveal surrounding logic, we introduce a closure-based subgraph extraction mechanism that reconstructs contextual neighborhoods around flagged paths.

Prior works extract h-hop enclosing subgraphs centered on individual nodes or edges~\cite{cai2021structural,louis2022sampling}. However, these radius-based expansions often introduce structural noise by including weakly related nodes. This leads to bloated subgraphs that reduce task efficiency in explainable or LLM-based analysis. To overcome this, we propose the $k$-hop Enclosing Subgraph. Instead of expanding from a single node, the $k$-hop Enclosing Subgraph captures the full set of predecessors and successors for each node along the path, yielding compact, semantically precise subgraphs that preserve invocation context with minimal noise. To capture the context around execution path, we iteratively include all in/out neighbors of the path nodes, controlled by a hop parameter $k$. Additional constraints on node degrees and subgraph size ensure the extracted subgraph remains tractable.

Let $\mathcal{G}$ be an execution call tree and a root-to-leaf path $P=(v_{p_0},\ldots,v_{p_\ell})$ with edge sequence $\mathbf{e}(P)=(e_1,\ldots,e_\ell)$, $e_j=(v_{p_{j-1}},v_{p_j})$, the directed neighbor set of a node $v$ can be drawn as 
\begin{equation}
 N(v) := \{\,u \in \mathcal{V} \mid (u,v)\in\mathcal{E} \ \lor\ (v,u)\in\mathcal{E}\,\},   
\end{equation}
Starting from the path nodes, the $k$-hop closure is derived recursively as
\begin{align}
\mathrm{C}_0(P) &= \{\,v_{p_0},v_{p_1},\ldots,v_{p_\ell}\}, \\
\mathrm{C}_{k}(P) &= \mathrm{C}_{k-1}(P) \cup \bigcup_{v\in \mathrm{C}_{k-1}(P)} N(v), \quad k \ge 1.
\end{align}

\begin{definition}[$k$-hop Enclosing Subgraph]
Based on the above recursive formula, the enclosing subgraph is defined as 
\begin{equation}
\mathrm{S}_k(P) = \mathcal{G}\big[\,\mathrm{C}_k(P)\,\big] = (\mathcal{V}_P,\mathcal{E}_P),
\end{equation}
where $\mathcal{V}_P=\mathrm{C}_k(P)$ and $\mathcal{E}_P=\{(u,w)\in\mathcal{E}\mid u,w\in\mathcal{V}_P\}$.

\end{definition}

Intuitively, $\mathrm{S}_k(P)$ starts from the path nodes ($k=0$) and expands outward up to $k$ hops. This prevents overly dense subgraphs and ensures consistent context extraction. Increasing $k$ may reveal higher-order dependencies but also risks introducing weakly related nodes. In~\autoref{sec:rq3} we empirically evaluate this trade-off.

Finally, we feed the outputs of our preceding modules into the LLM for report generation. Specifically, we pass the contract creation relations collected from the Detector, the corresponding contract bytecode extracted by the Extractor, and the enclosing subgraph of relevant execution paths obtained from the Analyzer into LLM as structured inputs. These inputs jointly capture both the structural context of contract interactions and the semantic details of their implementation, enabling the LLM to reason about the attack execution flow and summarize it into a comprehensive incident report. We also transmit the balance changes of each address before and after each transaction, which is obtained from the local node. The prompt is shown in~\autoref{fig:prompt_report} in Appendix.

\section{Experimental Evaluation}
\label{sec:eval}

In this section, we focus on the evaluation of TraceLLM. We will introduce the setup of experiments and show the evaluation results.

\subsection{Experimental Setup}
\label{sec:Implementation}
We use the large language model Gemini 2.0 Flash provided by Google via the Openrouter API \texttt{gemini-2.0-flash-001}. Default configurations were adopted, with temperature set to 0.7, top-p to 1, and a maximum response length of 2000. An Erigon full node was deployed to synchronize blocks and transactions with trace information. All experiments were executed in a Docker environment on Ubuntu 22.04, running an Intel Xeon 2.2 GHz processor with 64 GB RAM.

\subsection{Evaluation}

In this work, we aim to answer the following research questions (RQs):

\begin{itemize}
\item RQ1: (\textbf{Report Generation}) How accurately can TraceLLM generate security reports?
\item RQ2: (\textbf{Generalizability}) Can TraceLLM correctly identify attack methods across a larger set of real-world security events?
\item RQ3: (\textbf{Module Performance}) How well does each key module in TraceLLM perform?
\end{itemize}

\vspace{-1mm}
\spara{Methodology} To address RQ1 and RQ2, a dataset is constructed by aggregating all contract-vulnerability incidents reported by SlowMist~\cite{slowmist} from 2023 to 2025, supplemented with rug pull and private key leakage cases. Incidents lacking expert reports or sufficient address/time information are removed, yielding 148 cases with well-defined time ranges and addresses. Among these, 27 events with credible expert reports are selected to validate RQ1. Expert reports are treated as the ground truth. Reports generated by TraceLLM are evaluated using processed traces and suspicious execution paths in conjunction with decompiled contract code. To the best of our knowledge, this constitutes the first approach that automatically produces diagnostic reports for blockchain security. For comparison, a framework is constructed to emulate the methodology followed by experts when drafting reports. This pipeline integrates multiple code vulnerability analysis with raw traces, and the resulting reports serve as proxy baselines. To further answer RQ2, TraceLLM is applied to the remaining events, and report accuracy is assessed through expert judgment.

To answer RQ3, we evaluate the performance of the two most important modules of TraceLLM, which are Analyzer and Extractor. For Analyzer, we investigate whether Analyzer can effectively detect anomalous trace execution paths. We create the first anomaly trace dataset and split them into training and testing sets. We reconstruct call trees from transaction traces and label execution paths using human-written expert reports. We then compare our path scoring algorithm against representative statistical, machine learning, and neural network baselines. For Extractor, we evaluate the accuracy of our decompilation component by comparing reconstructed source code against ground-truth contracts. We select contracts from real-world security incidents and compare our approach to Panoramix, the decompiler currently used by Etherscan.

\begin{table}[!t]
\centering
\resizebox{\columnwidth}{!}{%
\begin{tabular}{l c c c c c}
\toprule
\textbf{Event} & \textbf{Att./Vit.} & \textbf{TraceLLM} & \textbf{Mythril} & \textbf{Slither} & \textbf{GPTScan} \\
\midrule
Conic\_1 & $\checkmark$ & $\checkmark$ & {\color{red}x} & $\checkmark$ & $\checkmark$ \\
Conic\_2  & $\checkmark$ & $\checkmark$ & {\color{red}x} & $\checkmark$ & $\checkmark$ \\
Aave  & {\color{red}x} & {\color{red}x} & {\color{red}x} & {\color{red}x} & {\color{red}x} \\
Vow & $\checkmark$ & $\checkmark$ & $\checkmark$ & $\checkmark$ & $\checkmark$ \\
Onyx Protocol & $\checkmark$ & $\checkmark$ & {\color{red}x} & {\color{red}x} & {\color{red}x} \\
Uwerx network & $\checkmark$ & $\checkmark$ & {\color{red}x} & {\color{red}x} & {\color{red}x} \\
Unibot & {\color{red}x} & {\color{red}x} & {\color{red}x} & {\color{red}x} & {\color{red}x} \\
Fire & $\checkmark$ & {\color{red}x} & {\color{red}x} & {\color{red}x} & {\color{red}x} \\
Onyx & $\checkmark$ & {\color{red}x} & {\color{red}x} & {\color{red}x} & {\color{red}x} \\
Sorra & $\checkmark$ & $\checkmark$ & {\color{red}x} & $\checkmark$ & $\checkmark$ \\
Aventa & $\checkmark$ & {\color{red}x} & {\color{red}x} & {\color{red}x} & {\color{red}x} \\
Mirage & $\checkmark$ & $\checkmark$ & {\color{red}x} & {\color{red}x} & {\color{red}x} \\
MEV Bot & $\checkmark$ & $\checkmark$ & {\color{red}x} & {\color{red}x} & {\color{red}x} \\
HopeLend & $\checkmark$ & {\color{red}x} & {\color{red}x} & {\color{red}x} & {\color{red}x} \\
Astrid & {\color{red}x} & {\color{red}x} & {\color{red}x} & {\color{red}x} & {\color{red}x} \\
pSeudoEth & $\checkmark$ & $\checkmark$ & {\color{red}x} & {\color{red}x} & $\checkmark$ \\
DePay & $\checkmark$ & $\checkmark$ & {\color{red}x} & {\color{red}x} & {\color{red}x} \\
Zunami & $\checkmark$ & $\checkmark$ & {\color{red}x} & {\color{red}x} & {\color{red}x} \\
Bybit & $\checkmark$ & $\checkmark$ & {\color{red}x} & $\checkmark$ & $\checkmark$ \\
Fake Memecoin & $\checkmark$ & $\checkmark$ & $\checkmark$ & $\checkmark$ & $\checkmark$ \\
Sleepless AI & $\checkmark$ & $\checkmark$ & $\checkmark$ & $\checkmark$ & $\checkmark$ \\
Ordinal Dex & $\checkmark$ & $\checkmark$ & $\checkmark$ & {\color{red}x} & {\color{red}x} \\
Peapods & $\checkmark$ & $\checkmark$ & {\color{red}x} & $\checkmark$ & $\checkmark$ \\
stoicDAO & $\checkmark$ & $\checkmark$ & {\color{red}x} & $\checkmark$ & $\checkmark$ \\
Abattoir of Zir & $\checkmark$ & $\checkmark$ & $\checkmark$ & {\color{red}x} & {\color{red}x} \\
Exzo Network & $\checkmark$ & $\checkmark$ & $\checkmark$ & $\checkmark$ & $\checkmark$ \\
Raft Protocol & {\color{red}x} & {\color{red}x} & {\color{red}x} & {\color{red}x} & {\color{red}x} \\
\bottomrule
\end{tabular}
}
\caption{Comparison of TraceLLM with proxy baselines.}
\label{tab:rq4}
\end{table}

\subsubsection{RQ1: Report Generation}
We evaluate TraceLLM on 27 real-world security incidents that occurred within the past two years, each accompanied by an expert-written security report, which we treat as ground truth. For each incident, we collect all relevant transactions on the day of the attack, covering a total of 201{,}593 blocks. Our goal is to assess whether TraceLLM can accurately recover attacker and victim addresses, as well as identify the vulnerable functions and attack methods. Since there is currently no existing framework that integrates both trace-level analysis and code-level vulnerability detection, we constructed comparison pipelines by augmenting traditional code analysis tools with raw trace data. Specifically, we selected Slither~\cite{feist2019slither}, Mythril~\cite{mythril}, and GPTScan~\cite{sun2024gptscan} as representative approaches from static analysis, symbolic execution, and LLM-based vulnerability detection, respectively. For each baseline, we analyze the source code of the victim contracts identified in expert reports, sending the resulting vulnerability reports along with the relevant raw traces to LLM, and ask the model to generate a security incident report by using the same prompt in~\autoref{fig:prompt_report} in Appendix. This design enables a comparison between TraceLLM and prior tools.

Table~\ref{tab:rq4} reports the results. TraceLLM achieves 85.19\% precision in recovering attacker and victim addresses and correctly identifies the vulnerable functions and attack methods in 19 out of 27 cases, reaching a precision of 70.37\%. By contrast, the best-performing baseline, GPTScan report with transactions and traces, only achieves 44.44\% precision, while Slither- and Mythril-based pipelines perform substantially worse (40.74\% and 22.22\%, respectively). We observe that compared to TraceLLM, proxy baselines often produce misleading interpretations. For example, Slither-based pipeline frequently misclassifies rug pulls as reentrancy, while Mythril-based pipeline tends to misidentify other contract vulnerabilities as rug pulls.

These findings highlight two key insights. First, TraceLLM represents the first attempt to systematically combine transaction-level trace semantics with code reasoning for security incident analysis, a capability not previously explored in blockchain security research. Second, even when we strengthen existing code analysis tools by supplementing them with trace information, TraceLLM consistently outperforms them, demonstrating its unique ability to bridge semantic gaps between execution behavior and contract logic in real-world attacks.

\begin{tcolorbox}[title=Answer to RQ1, colback=blue!5!white, colframe=blue!75!black]
TraceLLM can successfully identify attacker/victim addresses with 85.19\% precision and generate high-quality analysis reports. It correctly detects 70.37\% of vulnerable functions and attack methods relative to the ground truth assessed by human experts, significantly outperforming the proxy baselines.

\end{tcolorbox}

\subsubsection{RQ2: Generalizability}
To evaluate the generalizability of TraceLLM across additional real-world incidents, we generate the analysis report for the remaining 121 security events. Expert analysis is performed to assess whether TraceLLM accurately identified attacker and victim addresses and successfully distinguished attack methods. Among these events, 79 reports are correctly produced. Combining with the results from RQ1, TraceLLM achieves 82.43\% precision of attacker/victim detection and an overall precision of 66.22\% across 148 real-world security events.

\begin{figure}[t]
    \centering
    \includegraphics[width=1\linewidth]{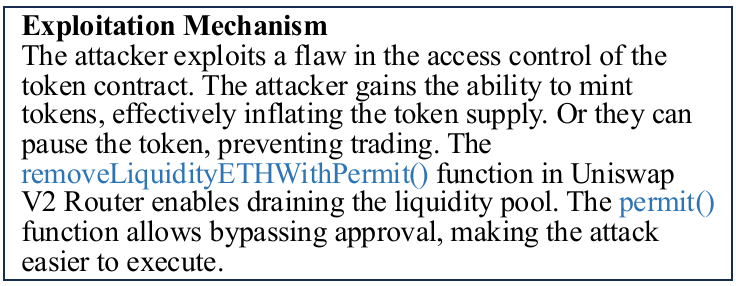}
    \caption{The exploitation mechanism of the PlayDapp hack.}
    \label{fig:attack_execution1}
\end{figure}

\begin{figure}[t]
    \centering
    \includegraphics[width=1\linewidth]{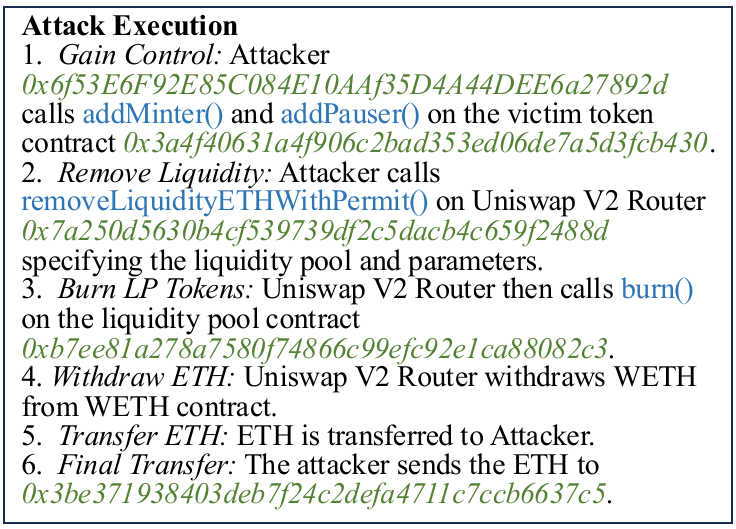}
    \caption{The attack execution of the PlayDapp hack.}
    \label{fig:attack_execution2}
\end{figure}

We use a representative event to illustrate the report showcasing TraceLLM. In mid-February 2024, PlayDapp was hacked, resulting in a loss of \$290 million. According to records in the SlowMist historical database, the attack method was private key leakage, and the attacker’s address was added as a token miner~\cite{PlayDapp}.~\autoref{fig:attack_execution1} illustrates TraceLLM’s exploitation mechanism for analyzing this incident. The cause of the attack and the vulnerable functions are accurately identified. TraceLLM also extract attacker and victim addresses and reconstruct the execution flow of the exploit, as shown in~\autoref{fig:attack_execution2}. Expert evaluation confirm that the report provides an accurate analysis of the PlayDapp incident while remaining consistent with the SlowMist event library, whose descriptions are more ambiguous than TraceLLM’s analysis.

\begin{tcolorbox}[title=Answer to RQ2, colback=blue!5!white, colframe=blue!75!black]
TraceLLM exhibits strong generalizability in analyzing real-world security incidents. We extend our evaluation from 27 to 148 real-world incidents, and TraceLLM achieves 82.43\% precision of attacker/victim detection and 66.22\% precision in the exploitation mechanism and attack execution analysis in reports.
\end{tcolorbox}

\subsubsection{RQ3: Module Performance}
\label{sec:rq3}
After demonstrating the overall report generation capability and generalizability of TraceLLM, attention is shifted to the internal modules influencing report quality. Within TraceLLM, the most critical factors are whether the anomaly execution path is correctly identified in the Analyzer and whether the Extractor produces accurate decompiled code. A detailed analysis of these two modules is conducted. 

\vspace{-1mm}
\spara{The performance of the Analyzer} We evaluated our method on 15 real-world blockchain security incidents with human-written expert reports to measure the accuracy of anomaly execution path identification. For each incident, we constructed call trees from all relevant transaction traces, extracted execution paths via DFS, and labeled them by matching victim contracts, vulnerable functions, and attacker addresses from the reports. This yielded 11{,}228 unique paths, of which 1{,}530 were labeled as primary attack paths. To the best of our knowledge, this is the first publicly available dataset that systematically identifies anomalous execution paths within blockchain transaction traces, rather than only detecting anomalous transactions at a coarse granularity. By releasing this dataset and its ground-truth annotations, we provide the first benchmark for evaluating anomaly trace detection methods in the blockchain security domain.

We compare our path scoring algorithm against representative statistical, machine learning, and neural network baselines. Specifically, for statistical methods, we include (1) a semantic-based univariate statistical scoring (Semantic) according to~\autoref{eq:semantic}, and (2) the Priority Score (Score) widely used in backward and forward causality tracking for system security~\cite{liu2018towards}. For machine learning, we select Random Forest (RF) and XGBoost, two well-established models for anomaly detection due to their robustness to high-dimensional sparse features and ability to capture non-linear patterns~\cite{primartha2017anomaly,li2023survey}. For neural networks, we use MLP, GIN and SAGE, three representative graph neural network (GNN) architectures capable of modeling structural dependencies in call trees~\cite{wang2021decoupling, sajadmanesh2023gap, xu2018powerful}. To our knowledge, this is also the first work to systematically evaluate such a diverse set of baselines on anomaly trace detection in blockchain systems.

Following a Leave-One-Group-Out(LOGO) evaluation strategy, we rank execution paths in each method by their anomaly score or predicted probability and select the top 20 ranked paths for each incident. Since our goal is to maximize the number of ground-truth attack paths exposed to the LLM, we focus on recall, defined as $\text{recall} = \frac{\#\text{Hit}}{\#\text{Hit} + \text{FN}}$, where \#Hit denotes the number of ground-truth attack paths ranked in the top 20, and FN is the number of ground-truth paths missed.

\begin{figure}[!bpt]
    \centering
    \includegraphics[width=0.99\linewidth]{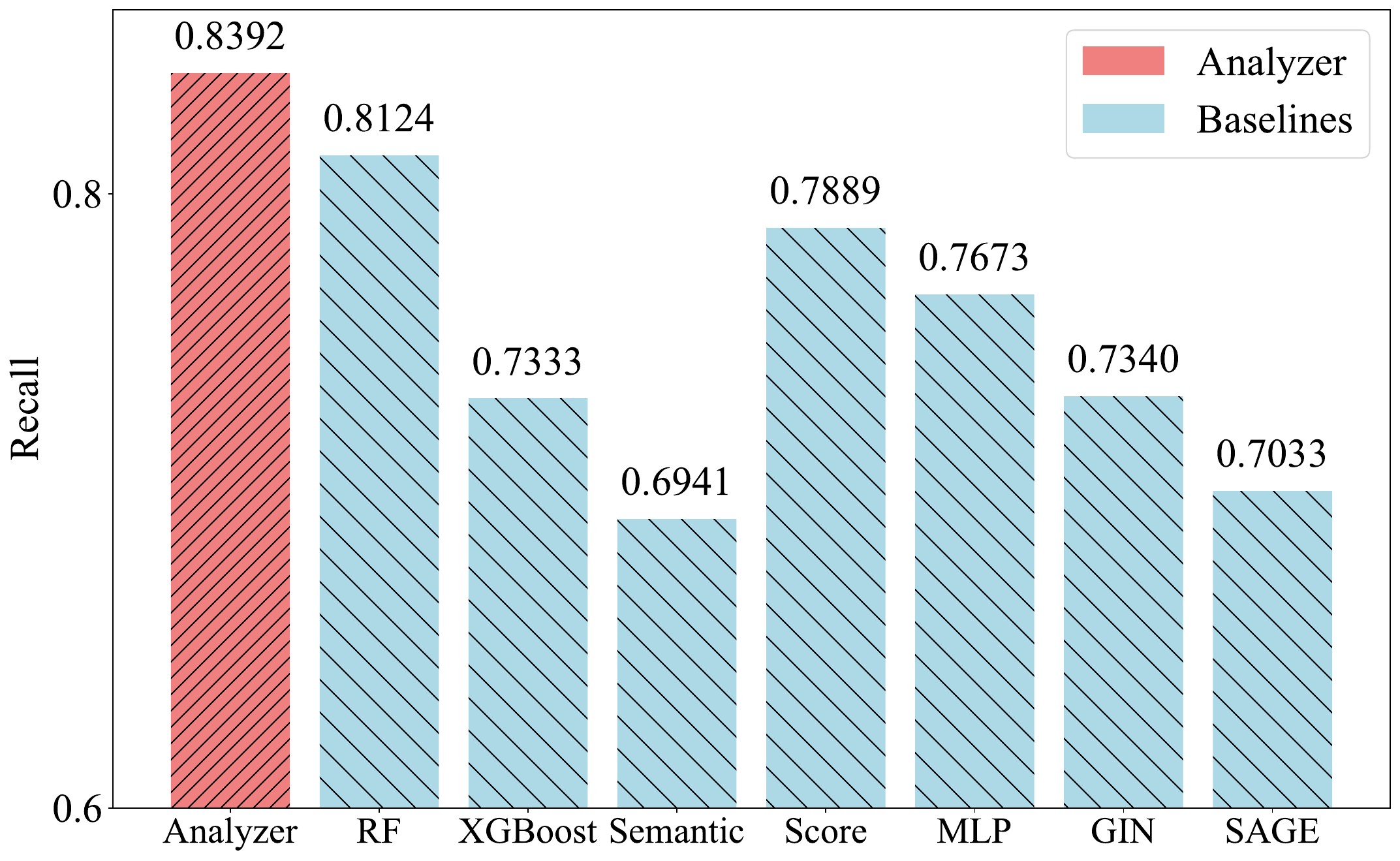}
    \caption{The recall of Analyzer in TraceLLM.}
    \label{fig:15logo}
\end{figure}

\autoref{fig:15logo} shows the average recall across 15 LOGO folds. Our solution achieves the highest recall (0.8392), outperforming all baselines, including Random Forest (0.8124) and Priority Score (0.7889). GIN (0.7340) and SAGE (0.7033) lag behind, suggesting that generic GNN architectures struggle to capture the semantic and hierarchical features of EVM call trees in this anomaly detection setting. The semantic-based univariate statistical method performs worst (0.6941), highlighting the limitations of ignoring multi-path contextual dependencies. These results demonstrate that our approach effectively integrates semantic and structural features of execution paths, yielding more accurate anomaly trace detection than traditional anomaly path detection baselines and establishing the first reproducible benchmark for this problem.

We also evaluate what the best value of $k$ is in~\autoref{sec:k-hop}. As ground truth, we adopt attacker, victim, and helper addresses extracted from expert incident reports. We then measure model performance under different closure depths ($k$-hop neighborhoods from 0 to 5). For each setting, we report the average token consumption and the precision of LLM-based predictions, where precision is defined as $\frac{\text{TP}}{(\text{TP}+\text{FP})}$, with true positives being correctly identified attacker or victim addresses. The results in~\autoref{tab:rq3} show that with $k=1$, the model correctly predicts attacker and victim addresses in 80\% of the 15 events, representing a 6.6\% improvement over the baseline without closure-based subgraph extraction. Increasing the closure depth to $k=2$ or $k=3$ raises token consumption by 75\% but does not yield further accuracy gains. At $k=4$ and $k=5$, token usage continues to grow while prediction accuracy declines. These findings suggest that $k=1$ achieves the best trade-off between accuracy and efficiency.

\begin{table}[!t]
\centering
\resizebox{\columnwidth}{!}{%
\begin{tabular}{c c c}
\toprule
\textbf{$k$-hop} & \textbf{Precision (\%)} & \textbf{Avg. Tokens Consumed} \\
\midrule
0 & 73.3 & 9{,}067.6 \\
1 & 80.0 & 62{,}444.8 \\
2 & 80.0 & 138{,}609.5 \\
3 & 80.0 & 196{,}163.2 \\
4 & 73.3 & 229{,}734.1 \\
5 & 80.0 & 248{,}614.5 \\
\bottomrule
\end{tabular}
}
\caption{Impact of different $k$-hop settings on address identification.}
\label{tab:rq3}
\end{table}

\vspace{-1mm}
\spara{The performance of the Extractor} We evaluated the accuracy of our Extractor module by comparing reconstructed source code against the ground-truth contracts. We randomly sampled 100 contracts from real-world security incidents, each with publicly available source code, and used the published source code as the ground truth. We then decompiled the corresponding on-chain bytecode using both our framework and Panoramix, a widely used Ethereum decompiler that serves as a strong baseline. To assess equivalence between decompiled code and ground truth, we adopt three top-ranking large language models: OpenAI-o1, Claude 3.5, and DeepSeek-R1. The prompt is provided in~\autoref{fig:prompt_consistent} in Appendix. For each contract, the models are given explanations derived from both the ground-truth source code and the decompiled code, and independently judge whether the two are consistent. A contract is marked as correctly decompiled only if all three models agree on consistency.

% \begin{figure}[!bpt]
%     \centering
%     \includegraphics[width=1\linewidth]{figures/consistent_prompt.pdf}
%     \caption{Prompt for consistent judging.}
%     \label{fig:prompt_consistent}
% \end{figure}

\begin{figure}[!bpt]
    \centering
    \includegraphics[width=0.47\textwidth]{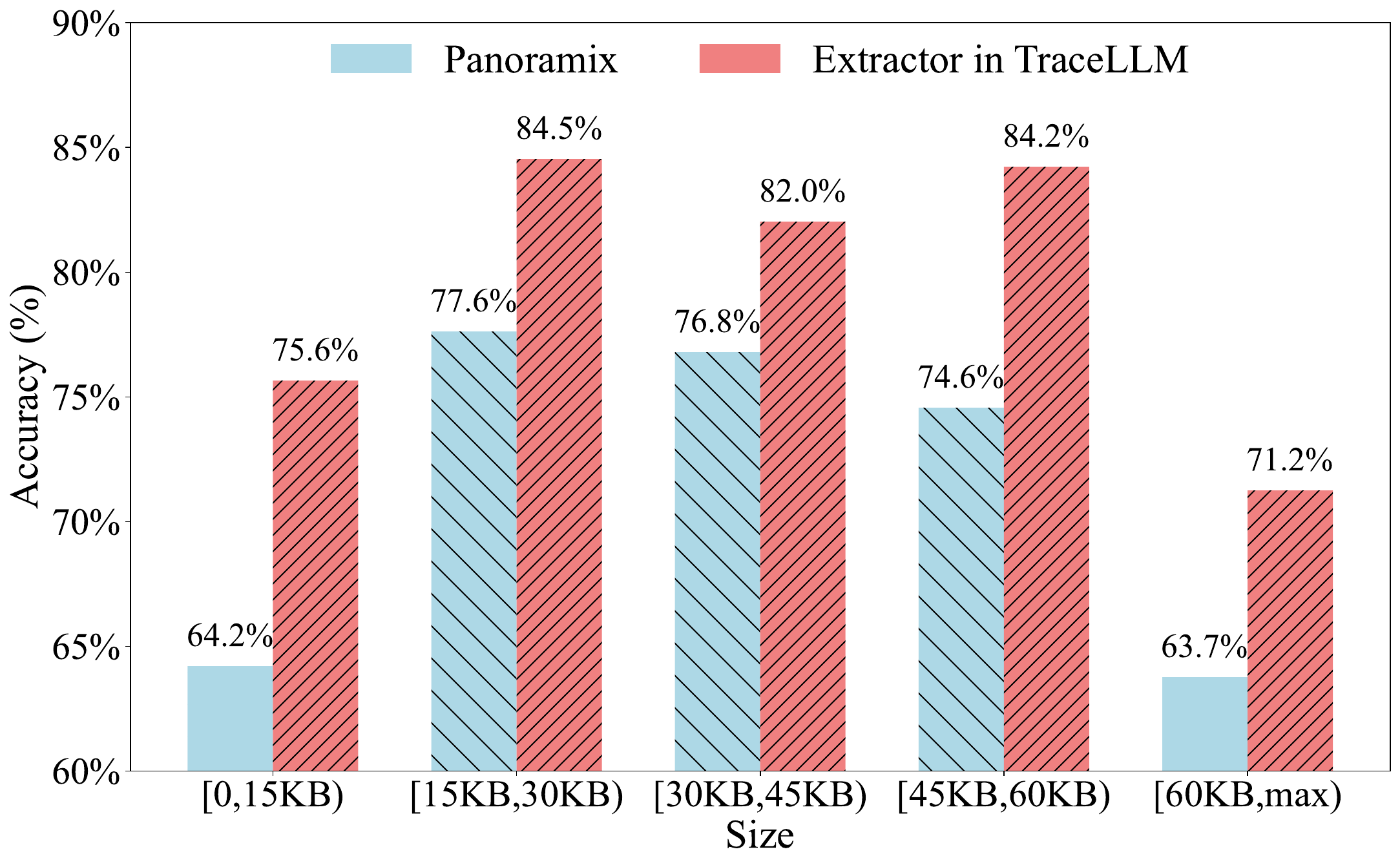}
    \caption{Accuracy of Extractor in TraceLLM.}
    \label{fig:comparison_}
\end{figure}

~\autoref{fig:comparison_} presents the accuracy comparison across different contract size ranges. The red bars represent our framework, and the blue bars correspond to Panoramix. Overall, Panoramix achieves 70.25\% average accuracy, while our framework improves upon this by 8.52\%. Notably, in the 0–15 KB and >60 KB contract size ranges, both approaches perform less effectively. This can be attributed to the fact that very small contracts often employ highly optimized, condensed bytecode with minimal structure, while very large contracts tend to contain complex, deeply nested control flows and large libraries, both of which challenge decompilation accuracy. Nevertheless, even in these challenging regimes, our framework consistently outperforms Panoramix, demonstrating its robustness across diverse contract sizes.

\begin{tcolorbox}[title=Answer to RQ3, colback=blue!5!white, colframe=blue!75!black]
The Analyzer module detects 83.92\% of anomaly execution paths, outperforming other anomaly detection methods. It achieves the highest accuracy with the lowest token consumption at k=1. The Extractor module correctly decompiles 78.77\% of unverified contracts, exceeding the most popular method by 8.52\%.
\end{tcolorbox}

\section{Discussion}
\label{sec:discussion}
In this section, we focus on the extensibility and limitations of TraceLLM. 

\subsection{Extension to Other Blockchains}
A notable strength of our framework lies in its extensibility beyond the Ethereum mainnet. Since our pipeline fundamentally relies on standardized EVM execution semantics, the methodology is readily applicable to other EVM-compatible blockchains. This generality is further reinforced by recent infrastructure advances: for instance, the upgraded Etherscan V2 API now supports unified queries across more than 50 chains with a single key~\cite{etherscanCreator}, enabling streamlined access to contract code and execution traces across heterogeneous ecosystems. Beyond cross-chain extensibility, our design also demonstrates conceptual extensibility in how LLM can jointly reason over smart contract and transaction traces. This integration suggests a broader range of applications than anomaly-driven forensic analysis. For example, with suitable prompt adjustments and minor pipeline changes, the system can produce human-readable transaction reports for routine interactions. This enables users to understand a transaction's intent before signing, enhancing user-centric security and transparency in decentralized applications. Taken together, these dimensions of extensibility highlight both the technical adaptability and the broader applicability of our approach.

\subsection{Limitations}

Despite the promising results, our system still faces several inherent limitations. First, the accuracy of code understanding is constrained by the precision of current decompilation tools. For highly complex or deliberately obfuscated contracts, the recovered pseudo-code often fails to preserve critical semantics, which restricts the ability of our pipeline to reconstruct attacker logic. Second, large language models themselves impose scalability constraints: analyzing intricate contract interactions frequently requires long-range reasoning across multiple layers of function calls, which may exceed the context length or reasoning capability of state-of-the-art models. Finally, our trace-based anomaly detection is limited by the granularity of path selection. In practice, some vulnerabilities originate from deeply nested internal functions that are only exposed to external users after several layers of delegation. In such cases, the abnormal path flagged by the detector may not accurately capture the root cause of the incident, reducing the precision of the reports. Addressing these limitations would require advances in both program analysis techniques and model architectures, as well as new methods for selectively capturing deeper execution context.

\section{Related work}\label{sec:relate}
\vspace{-1mm}
\spara{LLM for Blockchain}
The study of LLM for blockchain has been thoroughly examined in prior academic research, revealing significant understandings of its prevailing dynamics and potential future progressions~\cite{wang2024smartinv, chen2023chatgpt, shou2024llm4fuzz, yang2024hyperion, jiang2024unearthing}. BlockGPT~\cite{gai2023blockchain} and ZipZap~\cite{hu2024zipzap} were proposed to detect anomalous activities. In a similar manner, Sun et al.~\cite{sun2024gptscan} employed ChatGPT to identify vulnerabilities in smart contracts. The studies in~\cite{david2023you, wei2024llm} investigated the use of LLM in blockchain auditing. Liu et al~\cite{liu2024propertygpt} proposed PropertyGPT embedding network properties to detect more vulnerables. Also, LLM was utilized for contract auditing in~\cite{ma2025combining}. 
To the best of our knowledge, we are the first to integrate transaction execution traces and contract codes, employing LLM to automate the diagnosis of security events.

\spara{On-chain Analysis}
Literatures have explored numerical analysis of blockchain data~\cite{taverna2023snapping, yousaf2019tracing, zhang2023your, torres2021frontrunner, huang2024two, mclaughlin2023large}. Wang et al. ~\cite{wang2024understanding} presented a new fuzzing tool which can detect asymmetric DoS bugs, while Wang et al.~\cite{wang2025private} validated that the existing consensus protocol in Ethereum tends to monopolistic conditions. The study conducted in~\cite{wang2023automated} proposed two types of security properties to detect various types of finance-related vulnerabilities. Additionally, a taint analyzer was designed based on static EVM opcode simulation in~\cite{sun2024all}, identifiying more vulnerable contracts. Yaish et al.~\cite{yaish2024speculative} introduced 3 types of attack transactions based on Turing-complete contracts. Sun et al.~\cite{sun2023panda} first analyzed the semantics of Algorand smart contracts and find 9 types of generic vulnerabilities. Qin et al.~\cite{qin2023blockchain} uncover blockchain imitation game and the implications, and Miedema et al~\cite{miedema2023mixed} explored the mixing servise in bitcoin. Evaluation finished in~\cite{zhou2020ever} examined real attacks and defenses in smart contracts and revealed the consequences. 

\iffalse
A Mixed-Methods Study of Security Practices of Smart Contract Developers

Snapping Snap Sync: Practical Attacks on Go Ethereum Synchronising Nodes

Anatomy of a {High-Profile} Data Breach: Dissecting the Aftermath of a {Crypto-Wallet} Case

Total Eclipse of the Heart – Disrupting the InterPlanetary File System

EOSAFE: Security Analysis of EOSIO Smart Contracts

Evil Under the Sun: Understanding and Discovering Attacks on Ethereum Decentralized Applications

SmarTest: Effectively Hunting Vulnerable Transaction Sequences in Smart Contracts through Language Model-Guided Symbolic Execution

Smart Contract Vulnerabilities: Vulnerable Does Not Imply Exploited

EVMPatch: Timely and Automated Patching of Ethereum Smart Contracts

ETHBMC: A Bounded Model Checker for Smart Contracts

TXSPECTOR: Uncovering Attacks in Ethereum from Transactions

BlockSci: Design and applications of a blockchain analysis platform

The art of the scam: Demystifying honeypots in ethereum smart contracts
\fi
%\vspace{-0.1in}
\section{Conclusion}
\label{sec:con}

In this paper, we present TraceLLM, an LLM-driven framework that links Ethereum execution traces with contract code to automate post-incident security analysis. Unlike prior methods limited to either transaction detection or code analysis, TraceLLM combines trace anomaly detection and contract semantics to infer attacker and victim addresses, identify vulnerable functions, and uncover explicit attack mechanisms. Its modular pipeline, consisting of the Parser, Detector, Extractor, and Analyzer, tackles challenges such as proxy-based indirections, high-volume traces, and unverified contracts, while producing human-readable security reports. Extensive experiments on real-world incidents demonstrate that TraceLLM provides accurate and comprehensive forensic insights compared to existing tools. Extensive evaluations on real-world incidents show that TraceLLM delivers accurate and comprehensive forensic insights, establishing the first reproducible benchmark for automated blockchain forensics and demonstrating its potential to enhance both the automation and reliability of security investigations.

\newpage
\bibliographystyle{ACM-Reference-Format}
\bibliography{ref}

\newpage
\appendix
\section{Prompts}
\label{sec:prompts}

\begin{figure}[!bpt]
    % \vspace*{-\topskip}
    \centering
    \includegraphics[width=1\linewidth]{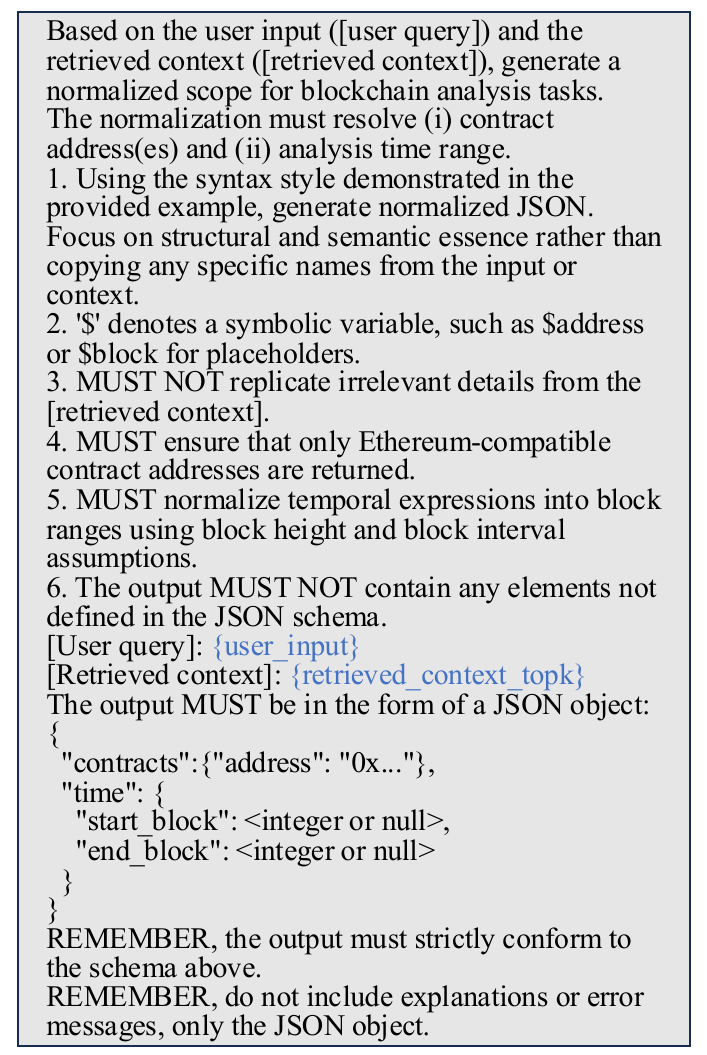}
    \caption{Generation prompt for query normalization.}
    \label{fig:one_shot_prompt}
\end{figure}

\begin{figure}[!bpt]
    \centering
    \includegraphics[width=1\linewidth]{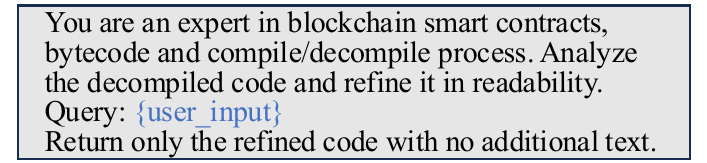}
    \caption{Prompt for code refining.}
    \label{fig:prompt_refine}
\end{figure}

\begin{figure}[!bpt]
    \centering
    \includegraphics[width=1\linewidth]{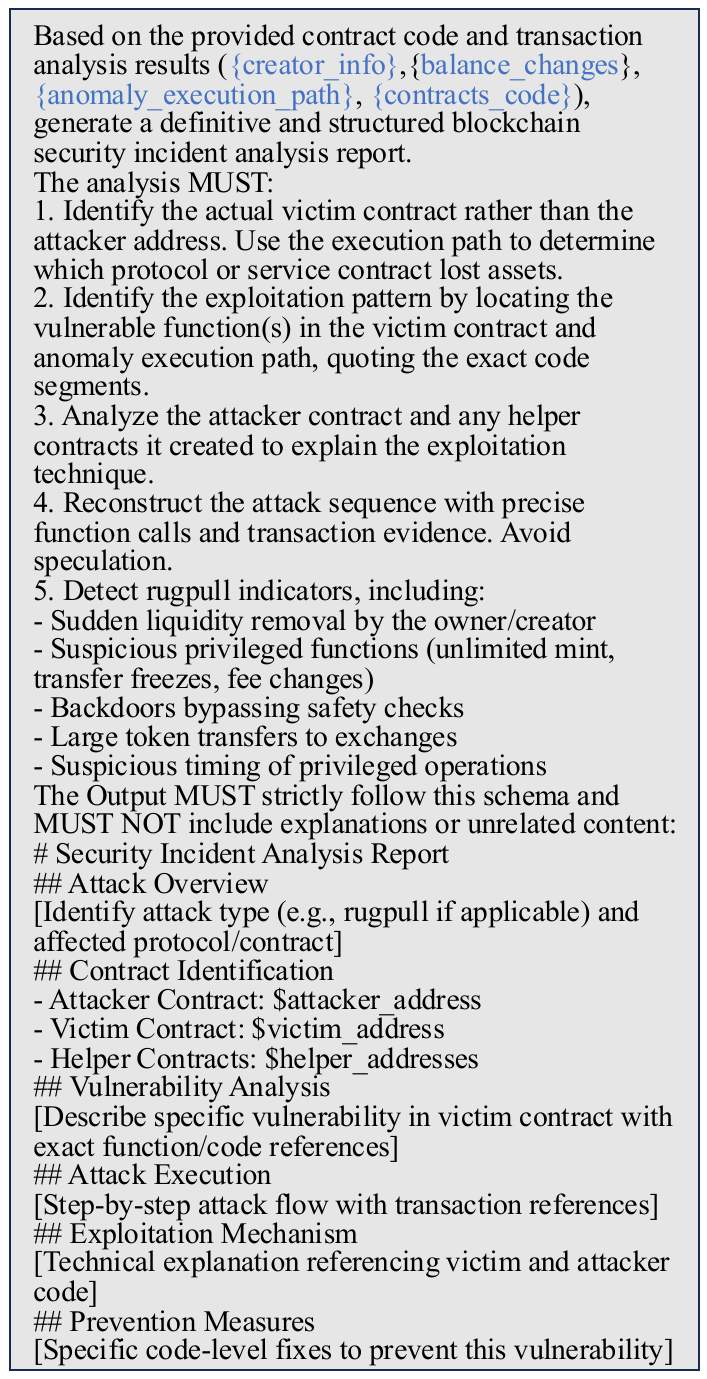}
    \caption{Prompt for report generation.}
    \label{fig:prompt_report}
\end{figure}

\begin{figure}[!bpt]
    \centering
    \includegraphics[width=1\linewidth]{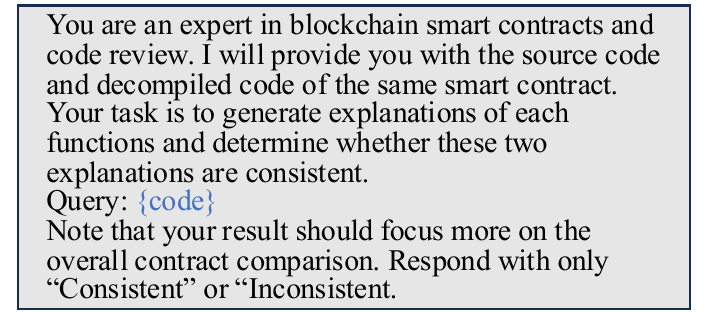}
    \caption{Prompt for consistent judging.}
    \label{fig:prompt_consistent}
\end{figure}

This section provides prompt used in TraceLLM. 
~\autoref{fig:one_shot_prompt} provides the prompt for generating query normalization.
~\autoref{fig:prompt_refine} provides the prompt for code refine.
~\autoref{fig:prompt_report} provides the prompt for report generation. 
~\autoref{fig:prompt_consistent} provides the prompt for explanation consistent judging. 

\end{document}